\documentstyle[aps,twocolumn,epsf]{revtex}
\begin{document}
\draft
 \title{
Superconductivity in correlated disordered two-dimensional electron gas
}
\author{J.S. Thakur and D. Neilson}
\address{
School of Physics, The University of New South Wales, Sydney 2052,
Australia.
\\[3pt]
(To appear in Phys.\ Rev.\ B )
\  \\ \medskip}\author{\small\parbox{14cm}{\small
We calculate the dynamic effective electron-electron interaction
potential for a low density disordered two-dimensional electron gas.
The disordered response function is used to calculate the effective
potential where the scattering rate is taken from typical mobilities
from recent experiments.  We investigate the development of an
effective attractive pair potential for both disordered and disorder
free systems with correlations determined from existing numerical
simulation data.  The effect of disorder and correlations on the
superconducting critical temperature $T_c$ is discussed.
 \\[3pt]{PACS numbers:
 73.20.Dx,71.30.+h,71.55.-i} 
}}\address{} \maketitle

\narrowtext

Recent experiments have revealed a well defined metal-insulator transition
(MIT) in 2D electron systems. The MIT has been observed in number of
different systems, Si MOSFET \cite{Kravchenko}, GaAs/AlGaAs \cite{SHG} and
p-SiGe \cite{MDH} for both electrons and holes. A common feature near the
transition is that the average electron-electron interaction energy is an
order of magnitude larger than the Fermi energy indicating that the
transition is driven by the strong Coulomb interactions. 

The transition from the metallic phase has been found to occur at
widely different levels of disorder and mobility.  In Si MOSFETs the
disorder at the transition is strong \cite{KRAV} while in high mobility
SiGe samples \cite{MDH} the disorder is very small.

Another feature of the MIT is that the resistivity scales with
temperature\cite {DPF} near the transition according to
$\rho(T,n_s)=\rho(T/T_o)$. The single parameter $T_o$ is a function of
$\delta_n=(n_s-n_c)/n_c$ on both the metallic and insulating sides of
the transition. $n_s$ is the electron density and $n_c$ the critical
density at the transition. As a result of this scaling all the data can
be collapsed into two branches, an upper one for the metallic phase and
a lower one for the insulator. The scaling behavior of the resistivity
is also observed with external electric field \cite{KS2}. Such scaling
is characteristic of a true phase transition. It is similar to the
superconductor-insulator phase transition \cite {STB} in thin
disordered films and the quantum Hall liquid-insulator phase
transition.  Independent measurements of the I-V characteristics at low
temperature and the magnetic susceptibility are required before the
existence of a superconducting phase can be confirmed. 

Thakur and Neilson \cite{TD1} have proposed that the insulating phase
of the low density system with weak disorder would be a frozen solid
phase with liquid-like short-range order.  They showed such a phase
could exist for electron densities 
corresponding to 
$r_s\agt$7 at impurity levels which
were in good agreement with the observations of the MIT.

The metallic phase near the transition is not a conventional metallic
state for which the Coulomb interaction energy is quite small compared
to Fermi energy. If that had been the case properties could have been
determined perturbatively in powers of the interaction potential.
However, near the MIT the electrons are strongly correlated on both the
metallic and insulating sides.  The scaling of the conductivity on both
sides with a common parameter $T_o$ indicates that the mechanisms for
electrical transport in the insulating and metallic phases are
related.

A number of proposals have been made about the nature of the metallic
phase. One suggestion is superconducting pairing of the electrons
\cite {PWI} with the pairing mediated by the dynamic correlation hole
surrounding each electron.  At very low densities the correlation hole
and the electron move rigidly as a single unit, and in the extreme case
of the Wigner crystal the correlation hole and the electron form the
Wigner-Seitz cell.  In the conducting phase at comparatively higher
densities than this the correlation hole is not rigidly bound to its
electron and on a time scale of the inverse of the plasmon frequency
there is the possibility of relative movement leading to partial
dissociation.  Over this time scale the partially vacated correlation
hole can attract another electron.  This attraction can lead to
superconducting pairing of the electrons.  The existence of a strong
correlation hole and weakly damped plasmons are both necessary for this
mechanism.  When the correlation hole is weak or disorder strongly
damps the plasmon there is no attraction.

In this paper we investigate the combined effect of disorder and
correlations on the superconductivity. We find that the
superconductivity persists at levels of disorder where the plasmon is  
damped out.  We also find that the superconducting transition
temperature is more sensitive to disorder when the electron correlations
are weaker, that is at higher electron densities.
 
To treat the disorder we introduce the generalized random phase response 
function 
\begin{equation}
\chi(q,\omega)=\frac{\chi^{(s)}(q, 
\omega)}{1+V(q)\left[1-G(q)\right]\chi^{(s)}(q,\omega)}\ .
\label{chiSTLSgamma} 
\end{equation} 
The local field factor $G(q)$ takes into account the correlations.
These make the effective electron-electron interaction weaker than 
the bare Coulomb potential $V(q)=2\pi e^2/{q\epsilon}$.
The effect of disorder is contained in the susceptibility
$\chi^{(s)}(q,\omega)$.  This is the particle conserving response
function of non-interacting electrons scattering from the disorder
\cite{Mermin},
\begin{equation}
\chi^{(s)}(q,\omega)=\frac{\chi^{(0)}(q,\omega+{\text{i}}\gamma)}
{1-\frac{{\text{i}}\gamma}{ \omega+{\text{i}}\gamma}
\left[1-\frac{\chi^{(0)}(q,{\omega+{\text{i}}\gamma})}
{\chi^{(0)}(q)}\right]}\ .
\label{chist(s)}
\end{equation} 
$\gamma$ is the scattering rate off the disorder.  
$\chi^{(0)}(q)$ is the static susceptibility for non-interacting
electrons without disorder.  $\chi^{(0)}(q,{\omega+{\text{i}}\gamma})$
is the dynamical susceptibility for non-interacting electrons
scattering off the disorder.
In the diffusive regime
$\lim_{\omega,q\rightarrow0}\chi^{(s)}(q,\omega)= (2m^\star/\pi
k_F\hbar^2)\ \{{{\cal{D}}q^2}/({{\cal{D}}q^2+\text{i}\omega})\}$, where
${\cal{D}}=v_F^2/\gamma$ is the diffusion constant. This vanishes as
$\gamma$ goes to infinity, representing the transition to the
insulating phase \cite{JLD}.

The scattering rate $\gamma$ could be calculated using the memory
function formalism for the density relaxation function and applying
mode coupling theory \cite{TN}. However, here we take $\gamma$ as a
parameter, the value of which is representative of the experimental
mobilities observed in recent studies. $\gamma$ is related to the
electron mobility $\mu$ by the Drude expression
$\mu=e/{m^\star\gamma}$.

The local field factor $G(q)$, taking into account electron
correlations, is calculated from numerical simulation data of the
static structure factor $S(q)$ \cite{Ceperley}. Introducing the
fluctuation dissipation theorem, $G(q)$ can be determined if we use the
generalized random phase approximation for $\chi(q,\omega)$
(Eq.\ \ref{chiSTLSgamma}).

As the density is lowered $G(q)$ develops a peak centered on
$q\simeq2.4k_F$.  When the height of the peak exceeds unity there is
the possibility that the denominator in the expression for $\chi(q)$
(Eq.\ \ref{chiSTLSgamma}) will pass through zero and in that case
$\chi(q)$ would diverge.  Such a divergence would probably indicate a
charge density wave (CDW) \cite{SNS}.  However we found that the peak
in $G(q)$ is not sufficiently high to offset the polarization which
decreases rapidly with $q$, and we detected no CDW divergence in
$\chi(q)$.

In Ref.\ \cite {PWI} the local field factor was approximated with
parameterized forms for parallel and antiparallel spin which had the
correct large $q$ asymptotic limit for $G(q)$ \cite{NI}.  We call this
the Iwamoto Local Field Approximation (ILFA).  The parameters are
expressed in terms of compressibility and spin-susceptibility sum rules
in the long wave length limit.  This approximation interpolates $G(q)$
for intermediate values of $q$.  For superconducting pairing the most
important correlations are those acting over distances determined by
the Fermi wave vector $k_F$.  Thus it is essential to determine the
correlations in this intermediate $q$ range accurately.  We compare
our results in the zero defect case with those obtained within the ILFA.

We consider electron densities such that only the lowest energy
sub-band is occupied.  The effect of correlations on the effective
electron-electron interaction
\begin{equation}
V_{eff}(q,\omega)=\frac{V(q)}{\epsilon(q,\omega)}=
V(q)[1+V(q)\chi(q,\omega)]
\label{chi(s)}
\end{equation}
is shown in Fig.\ 1.  We plot Re $V_{eff}(q,\omega$$=$$\epsilon_{F})$ for
electron density 
corresponding to 
$r_s=10$.  The depth and structure of the attractive region of
$V_{eff}(q,\epsilon_{F})$ is sensitive to correlations, as can be seen
from the comparison with the $V_{eff}(q,\epsilon_{F})$ calculated
within RPA (that is by setting $G(q)=0$).  
$V_{eff}(q,\epsilon_{F})$ in RPA is only attractive for $q/k_F\alt0.2$.
This attraction is caused by the plasmon resonance and is not relevant
here.  When correlations are included, an attractive region
develops over a wide range of wave-vectors.  One already sees this
effect within the ILFA (also shown), but the attraction there is much
weaker than with our full $G(q)$.  The deep well near $q/k_F\simeq2.4$
is the result of the peak in $G(q)$ exceeding unity.  This makes the
term $\{V(q)[1-G(q)]\}$ in Eq.\ \ref{chiSTLSgamma} attractive.

 \begin{figure}
 \epsfxsize=9.5cm
 \epsffile{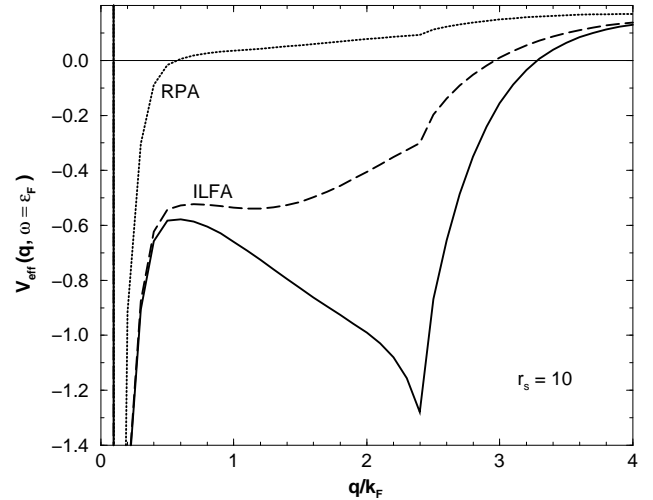}
  \caption[dummy3]{
Effective electron-electron interaction
$V_{eff}(q,\epsilon_{F})$ (solid line) for zero disorder $\gamma=0$.
Electron density 
corresponds to 
$r_s=10$. Dashed lines show $V_{eff}(q,\epsilon_{F})$ within RPA (no
electron correlations) and the ILFA (see text), as labeled.
 \label{Vq10}}
\end{figure}

In Fig.\ 2 we plot Re $V_{eff}(q,\omega$$=$$\epsilon_{F})$ for zero
disorder at various electron densities.  Increasing the density
decreases the correlations, and we see that the attractive part of
$V_{eff}(q,\epsilon_{F})$ weakens, especially around $q/k_F\simeq2.4$.  

In Fig.\ 3 we show the effect of disorder.
Re\ $V_{eff}(q,\omega$$=$$\epsilon_{F})$ is plotted for electron
densities corresponding to $r_s=10$ and $5$ at two values of the
disorder parameter $\gamma=0$ and $\gamma=2\epsilon_{F}$. 
This covers both the Si MOSFET samples
where typically 
$\gamma=2\epsilon_{F}$, and also the high mobility p-SiGe samples where
the value of $\gamma\ll\epsilon_{F}$.  In
the Figure we see that for $r_s=10$ the plasmon resonance around
$q/k_F\simeq0.3$ has been completely damped out by the time we reach
$\gamma=2\epsilon_{F}$.  Nevertheless the attractive well for
$q/k_F\agt2$ is not dramatically changed at this level of disorder.  In
contrast, at the higher electron density corresponding to $r_s=5$ the
disorder weakens the attractive potential.

 \begin{figure}
 \epsfxsize=9.5cm
 \epsffile{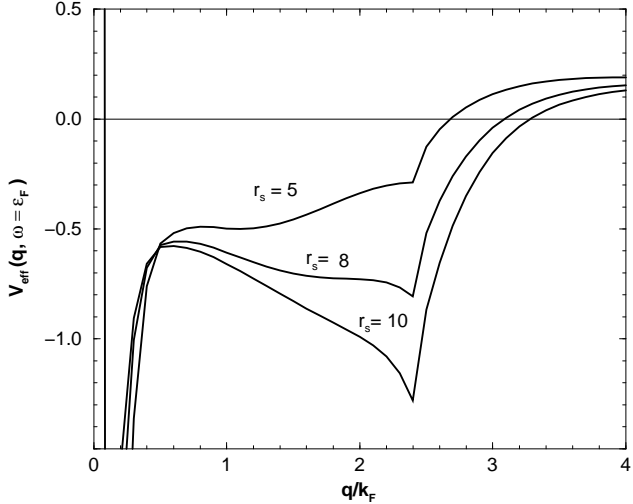}
  \caption[dummy3]{
 Effective electron-electron interaction $V_{eff}(q,\epsilon_{F})$ for 
zero disorder $\gamma=0$.  Curves are for different $r_s$ as labeled.
 \label{Vqrs}}
\end{figure}

We have calculated the critical superconducting temperature $T_c$ using
the Gor'kov equation for the gap function. In the Cooper channel the
irreducible interaction between two electrons at frequency $\omega$ and
$0$ is given by
\begin{equation}
F(\omega)=\frac{m^*}{2\pi}\int_0^{2\pi}\frac{\text 
d\theta}{2\pi}\int_0^\infty{\text
d\Omega}\frac{2}{\pi}\frac{|\omega|}{\Omega^2+\omega^2}V_{eff}(q,i\Omega). 
\label{FW} 
\end{equation}
Here, the wave vector $q$ is given in terms of the frequency $\omega$
as $q=\sqrt{p^2 + {p'}^{2} -2pp'\cos\theta}$, where
$p=\sqrt{2m^{\star}|\omega + \epsilon_{F}|}$ and $p'=k_{F}$.  The angle
between $\bbox p$ and $\bbox p'$ is $\theta$.  Under the weak coupling
approximation the critical temperature $T_c$ is given as \cite {Takada}
\begin{equation} 
T_{c}=1.134\epsilon_{F}\exp\left[-\frac{(1+\langle F\rangle)^2}
{\langle F^2\rangle -\lambda_0}\right] \ .
\label{Tc} 
\end{equation} 
The average $\langle F\rangle=
\int_{-\epsilon_{F}}^{\epsilon_{F}}{\text{d}}\omega(F(\omega)-F(0))/2\omega$.
We neglect any effect of disorder on $G(q)$ so $\lambda_0$=$\langle
F(0)\rangle$ is independent of disorder. The numerical simulation data
\cite{Ceperley} we have used is for disorder free systems.  If one used
a self-consistent calculation \cite{TN} in which the mutual dependence
of the disorder and the correlations were calculated then the
$\lambda_0$ would depend on disorder through the $G(q)$.

In Table 1 we show $T_c$ calculated for a range of $r_s$ and the
levels of disorder $\gamma=0$ and $2\epsilon_{F}$.  Disorder decreases
$T_c$ but the dependence is only significant at the higher densities:
by $r_s\agt10$ $T_c$ is insensitive to this disorder.  As $r_s$ decreases
there is an increase in $T_c$ until $r_{s}=6$$-$$7$.  Around this density
$T_c$ passes through a weak maximum and then starts to decrease.\\

 \begin{figure}
 \epsfxsize=8.0cm
 \epsffile{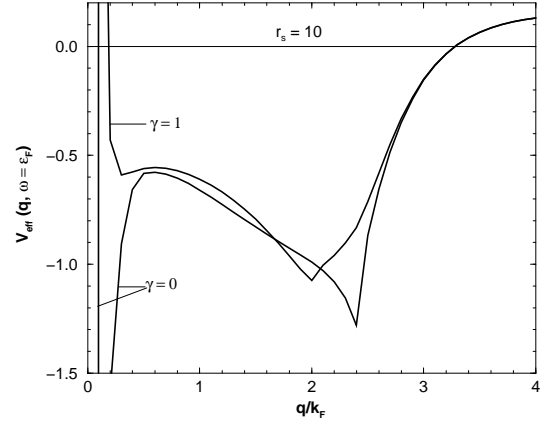}
 \epsfxsize=8.0cm
 \epsffile{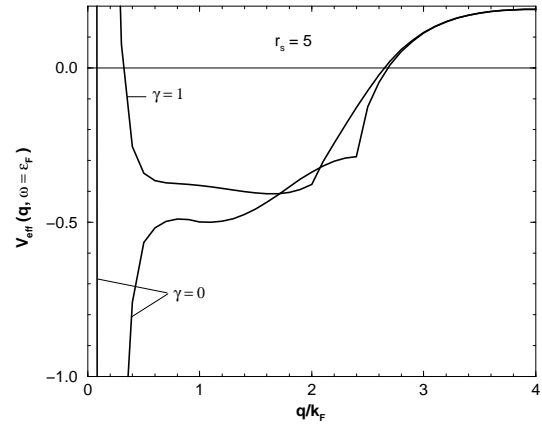}
  \caption[dummy3]{
 a. Effective electron-electron interaction
$V_{eff}(q,\epsilon_{F})$ for $r_s=10$.  The curves
are for two different values of the disorder parameter
$\gamma/2\epsilon_F$ as labeled.\\
 b. Same as (a) but for $r_s=5$.\\ 
 \label{Vqgam}}
\end{figure}

\[
\begin {array}{|c|c|c|}
\hline
\ &&\\
     r_s &    T_c\ (K)  &      T_c\ (K) \\
         &    \gamma=0  &      \gamma=2\epsilon_{F} \\
\ &&\\
\hline
   \mbox{\ \ \ \ \ \ \ \ \ } 5\mbox{\ \ \ \ \ \ \ \ \ }  & \mbox{\ \ \ \ \
\ \ \ \ } 0.62 \mbox{\ \ \
\ \ \ \ \ \ } &  \mbox{\ \ \ \ \ \ \ \ \ } 0.25 \mbox{\ \ \ \ \ \ \ \ \ } \\ 
   \ 6\  & \ 0.63 \ &  \ 0.43 \ \\ 
   \ 7\  & \ 0.58 \ &  \ 0.47 \ \\ 
   \ 8\  & \ 0.52 \ &  \ 0.46 \ \\ 
   \ 9\  & \ 0.46 \ &  \ 0.42 \ \\ 
   \ 10\ & \ 0.40 \ &  \ 0.38 \ \\ 
\hline
\end{array}
\]
\ \\
Table 1.  Superconducting transition temperature $T_c$ as a function of
$r_s$ for disorder levels $\gamma=0$ and $2\epsilon_{F}$ (see text).\\

The discrepancy between our $T_c$ and $T_c^{ILFA}$ calculated within
the ILFA is much larger at $r_s=5$ than at $r_s=10$.  At $r_s=5$ and
$\gamma=0$ ILFA gives $T_c^{ILFA}=0.27$ K which is a factor of two
smaller than our $T_c$, while at $r_s=10$ the $T_c^{ILFA}=0.34$ K.  The
maximum in $T_c^{ILFA}$ occurs near $r_s=8$.

In conclusion we find that electron correlations and the disorder
typical of that found in the current experimental samples has a
significant effect on the effective attractive interaction which acts
between the correlated electrons at low densities. Interestingly
while the quite high levels of disorder typical of those in the Si
MOSFET experiments are strong enough to completely damp out the
plasmon, the attractive interaction for $q/k_F\agt0.5$ persists because
of the static correlations.  The effective interaction is sensitive to
the details of the correlations and our results including the full
correlations significantly differ from the ILFA.  The estimated
superconducting transition temperature is sensitive to the details of
the correlations and, at higher densities, to the disorder.

\acknowledgements 

This work is supported by Australian Research Council Grant.

\end{document}